\documentclass[12pt]{article}

\begin{document}

\title{Can we really rule out the existence of antistars?}

\author{Luigi Foschini\\
\small{Institute TeSRE -- CNR, Via Gobetti 101, I--40129, Bologna(Italy)}\\
\small{email: \texttt{foschini@tesre.bo.cnr.it}}}

\date{}

\maketitle

\begin{abstract}
The existence of large domains of antimatter is still an open 
question. Some space mission and experiments in the near future are 
expected to give reasonable answer to that question.  
Meanwhile, we can try to search for other signatures of the presence 
of antimatter. This paper presents a discussion 
about possible effects of the close encounter of stars and antistars. 
It is known that the accretion power can be higher in presence of antimatter,
because of high--energy photons generated in annihilation, which 
result in a smaller radiation pressure. Therefore, the Eddington 
luminosity can reach higher values. Calculations of the gamma--ray flux
above 100~MeV, and comparison with recent data from satellites, show a good
agreement for certain sources. So, let's play some science--fiction!
\end{abstract}

\section{Introduction}
During last decades, high--energy astrophysics has dramatically 
changed, with the discovery of powerful gamma--ray sources. 
Gamma--ray bursts (GRB) represent one of the most fascinating 
mysteries of the nature.  Since their accidental discovery, in the 
late 1960s (Klebesadel et al. \cite{GRB1}), they puzzled
astrophysicists both for their enormous amount of energy released 
(about $10^{51}-10^{54}$ erg in a few seconds) and for their ``inner 
engine'', hidden from direct observations.  This started soon a 
plethora of theories, involving supernovae, neutron stars, antimatter 
effects, black holes, and more and more exotic options (for a review 
see Piran \cite{PIRAN1}, \cite{PIRAN2}, Rees \cite{REES}).

GRB are not the only source in the universe with large energy release;
for example, another exciting mystery is represented by active
galactic nuclei (AGN).  They are found to produce much more luminosity
than a typical galaxy, even $10^{4}$ times, in a tiny volume, less
than 1~pc$^{3}$ (for a review, Krolik \cite{KROLIK}).

It is worth noting that the energy release is too large to be due to
nuclear reactions and, if we want to avoid introducing new physics,
we can use only two sources of energy: gravitation and
matter--antimatter annihilation.
Actual theories consider gravitation as the main mechanism driving
these high--energy astrophysical sources, while matter--antimatter
annihilation has been neglected mainly because we have not detected
macroscopic bodies made of antimatter in the Solar System and its
neighbourhood.  Moreover, a well detailed review by Steigman
(\cite{STEIG1}) showed that there is no experimental evidence of
the existence of large amount of antimatter in the universe.  But
according to Steigman himself: ``On the other hand, failure to
construct consistent, symmetric cosmologies may indicate either a lack
of antimatter or a lack of imagination''.

Indeed, although experimental particle physics has showed the symmetry
between particles and antiparticles -- the creation of matter must
occurs simultaneously with the creation of antimatter --, it can be
unsatisfactory that current cosmologies consider a universe made of
matter only.  First attempts to construct a symmetric cosmology were
done by Klein and Alfv\'en during fifties and sixties (see Alfv\'en
\cite{ALF1}).  Later on, this argument was widely debated --
literature includes diametrically opposed points of view --, but there
are no sufficient data to select among several theories (Cohen et
al.~\cite{ANTI1}).  It is quite difficult to show the existence
of antimatter in the universe, but it is also difficult to show why we
have a universe made of matter only (Stecker~\cite{STECK}).

The recent developments of astrophysical technologies allowed the discovery
of several gamma--ray sources and it is time to ask ourselves if new discoveries
and measurements can overcome past objections to antimatter theories.

Among various theories, Alfv\'en~(\cite{ALF2}) proposed an annihilation
model for the energy source of quasi--stellar objects. It is a detailed paper
and we refer the reader to it, but there are some features that Alfv\'en
did not take into account and here we would like to examine them, in order
to extend, improve, and update the Alfv\`en theory.
Particularly, in this paper we investigate
some ideas about the close encounter of stars and antistars in order to
search possible signatures of antimatter. The notes exposed here do
\emph{not} want to propose a ``universal'' theory for astrophysical
gamma--rays sources. However, \emph{some} (not all) gamma sources
might be due to the close encounter of stars and antistars and
related phenomena.

\section{Matter or antimatter, that is the problem}
The main source of information about cosmic bodies is the
electromagnetic radiation they emit.  However, photons are not
subjected to CPT symmetry.  The interaction of antimatter with magnetic
fields is different from the interaction of matter, but the effects
(Zeeman; Faraday rotation) depend also on the sign of the magnetic
field, and therefore we cannot get any conclusions on the kind of
matter.

The best way to show the evidence of an antistar is surely the detection
of antinuclei with $Z>2$.
This was the purpose of the \emph{Alpha Magnetic Spectrometer} (AMS),
that was flown on the space shuttle in June 1998: first results indicate
no antihelium flux and pose an upper limit on the ratio of antihelium to
helium (Alcaraz et al.  \cite{AMS}).  Best results should be available
with a second experiment, that will have a sensitivity in antihelium search
of $10^{-9}$.

Another method was suggested by Cramer and Braithwaite~(\cite{CRAMER}). They
noted that the polarization of photons generated in fusion processes of stars is
different according to that there is matter or antimatter. While stars of matter
emit right--circularly polarized photons, antistars emit photon which are
left--circularly polarized.

Another way is to detect the products of annihilation: neutrinos,
$\gamma$--rays, and relativistic electron--positron pairs.  A typical
decay scheme for nucleon--antinucleon annihilation is
(Steigman~\cite{STEIG1}):

\[
N+\bar{N}\rightarrow\cases{\pi^{0}\rightarrow \gamma + \gamma \cr
\pi^{\pm}\rightarrow \mu^{\pm}+\nu_{\mu}(\bar{\nu_{\mu}})\cr}
\]

\noindent and:

\begin{displaymath}
	\mu^{\pm}\rightarrow
	e^{\pm}+\nu_{\mu}(\bar{\nu_{\mu}})+\nu_{e}(\bar{\nu_{e}})
\end{displaymath}

The spectra are in the range from several tens to several hundreds of
MeV, and with a peak around 100~MeV and a median energy around
250--300~MeV.

This is the most frequent annihilation process for free nucleons. The
generation of ``protonium'' (something similar to positronium, but
with protons and antiprotons) with following annihilation, has
a rate that is about $10^{-5}$ times smaller than the process
described above. Direct annihilation (that is
$p+\bar{p}\rightarrow\gamma+\gamma$) is quite rare,
with a rate about $10^{-7}$ times smaller  (Steigman~\cite{STEIG1}).

The annihilation of electrons and positrons gave two photons with
energy of 511~keV. Under certain environmental conditions,
specifically with gas temperature lower than about $5\cdot 10^{5}$~K
(Guessoum et al.~\cite{GUESS}), positronium can take place by mean
of radiative recombination. After some fraction of seconds, the particle
and the antiparticle annihilate with the production of two or three
photons, according to the state: two photons for the singlet and
three photons for the triplet. In this case, the spectrum presents a
continuum spectrum in the low energy side of the 511~keV line. If the
gas temperature is higher than $5\cdot 10^{5}$~K, the annihilation
takes place directly without the formation of positronium; therefore
we have only the 0.511~MeV line.

For example, OSSE observations of both galactic 511~keV line emission and
the three photon continuum components of the electron--positron annihilation
radiation showed the presence of a galactic region with positronium
(Kinzer et al.~\cite{OSSE1}, Purcell et al.~\cite{OSSE2}).
Therefore, the gas temperature in the galactic centre is lower than
$5\cdot 10^{5}$~K.

About half of the energy released in annihilation is carried away by
neutrinos, and some authors proposed to search there signatures for
antimatter (see Steigman~\cite{STEIG1}, Barnes et al.~\cite{BARNES}).
However, it is well known how difficult is to detect netutrinos.

Actually there are still large uncertainties in measurements: the INTEGRAL
satellite, for example, that will be launched on April 2002, will provide
for the first time fine gamma--ray spectroscopy. In this situation, there are
still several questions without answers of reasonable certainty and the
antimatter in the universe cannot be completely ruled out.
So we can speculate about possible signatures of its presence.

\section{Evaporation and ejection from a cluster}
Khlopov~(\cite{AMS2}) showed that it is possible that there
were isolated domains of antimatter, but they should be larger
than a globular cluster.
Antistars can escape from an anticluster in two ways (Binney
and Tremaine \cite{GALA}): ejection, that is the antistar gains the
escape speed by a single close encounter with another antistar in the
cluster; evaporation, that is a gradual increase of the speed from
several distant encounters.  The first case is negligible, when
compared to the second one, but the rate of evaporation is more
complicated to calculate.  A crude estimation is given in Binney and
Tremaine (\cite{GALA}), that put $t_{\mathrm{ev}}\approx 100
t_{\mathrm{rh}}$, where $t_{\mathrm{rh}}$ is the mean relaxation
time. This means that during a time $t_{\mathrm{ev}}$ a significant
fraction of stars leave the cluster. For typical globular cluster
$t_{\mathrm{rh}}$ is in the range between $3\cdot 10^{7}$ to
$2\cdot 10^{10}$ years (Meylan and Heggie \cite{HEGGIE}).

Globular clusters date back to the formation of galaxies and they are
located within a large roughly spherical halo around galaxies.  As
known, if there were isolated domains of antimatter, they could have
survived in globular clusters (Khlopov \cite{AMS2}).  If a cluster
of antistars survived near a galaxy of matter, then it is possible
that an antistar evaporates from the cluster and goes into the galaxy,
causing the annihilation and gamma--ray emission.  The tidal forces of
the galaxy can increase the evaporation rate and once in the galactic
nucleus, there is a high probability of binary formation by tidal
capture and stellar merging or collision (Lee and Ostriker
\cite{LEE}).

Under these conditions, a fraction of close encounters could be
between stars and antistars.  This value depends on the mass fraction
of antimatter domains relative to the total baryon mass, which in turn
is strongly model dependent.  For the sake of the simplicity, let us
suppose that, at least, one globular cluster near a galaxy was made of
antimatter.  We consider a mean total number of globular cluster for
each galaxy of about $2\cdot 10^2$.  In addition, we have to evaluate
the rate of evaporation through escaping stars, that in its simplest
form can be written as (Binney and Tremaine \cite{GALA}):

\begin{equation}
\frac{dN}{dt} = -\frac{N}{t_\mathrm{ev}}
\label{e:evapo}
\end{equation}

If we assume $N=10^6$ and consider a time interval of one year, we
obtain, in the best case, that $N\approx 3\cdot 10^{-4}$ star for each
year and for each cluster.  The rate of evaporation of an antistar
from an anticluster is then $1.5\cdot 10^{-6}$~yr$^{-1}$.  We remind
that we assumed that at least one globular cluster near a galaxy was
made of antimatter.

It is worth noting that the rate of escape from a cluster is a
function of star mass (see Meylan and Heggie \cite{HEGGIE}) and
values obtained here are strongly dependent on simplifying
assumptions.  Numerical models are required to give better results;
here we want only to give a rough evaluation.

The next step is then to study what happens during the close encounter
of a star and an antistar.

\section{Close encounters}
Close encounters can lead to tidal capture and, in certain cases, also
to collisions and merging. These processes are driven by gravity and therefore
there is no difference when we are dealing with stars or antistars.

Let us consider the mass transfer in a binary
system of a star and an antistar, which involves the accretion power.
The standard model explains the radiation emission by using the
gravitation as a source of energy: the kinetic energy of the infalling
matter is converted into radiation. There are two ways to do the mass
transfer: stellar wind or accretion disk due to Roche--lobe overflow
(see Frank et al. \cite{FRANK}). When antimatter is present,
we have additional energy release from annihilation.

The main effect of this additional energy is to change the Eddington
luminosity, that is found by balancing the inward gravitational force
and the outward radiation pressure, deriving from the conversion of
kinetic energy and from annihilation. The standard formula of
Eddington luminosity is:

\begin{equation}
	L_{\mathrm{Edd}}=\frac{4\pi GMm_{\mathrm{p}}c}{\sigma_{\mathrm{T}}}
	\approx 1.3\cdot 10^{38}\frac{M}{M_{\odot}}~\mathrm{erg/s}
	\label{e:edd}
\end{equation}

\noindent where $G$ is the gravitation constant, $M$ is the mass of
the accreting star, $m_{\mathrm{p}}$ is the mass of the proton, and
$\sigma_{\mathrm{T}}$ is the Thompson cross section:

\begin{equation}
	\sigma_{\mathrm{T}}=\frac{8\pi}{3}r_{\mathrm{e}}^{2}
	\label{e:th}
\end{equation}

\noindent with $r_{\mathrm{e}}$ the electromagnetic radius of the
electron.

For luminosities greater than $L_{\mathrm{Edd}}$ the radiation
pressure stops the accretion process, because it exceeds the inward
gravitational force.

When dealing with accretion processes between a star and an antistar,
the radiation pressure becomes very important. Photons produced by
annihilation of electron--positron pairs have energy of 511~keV, that
is the upper limit of validity of the Thompson cross section.
Annihilation of heavier particles produce photons with energy $\hbar\omega$
greater than $m_{\mathrm{e}}c^{2}$, as seen in the above section.
Therefore, we must use proper quantum relativistic cross section,
that is given by Klein--Nishina formula (see, for example,
Longair \cite{LONGAIR}):

\begin{equation}
\sigma_{\mathrm{KN}}=\frac{\pi r_{\mathrm{e}}^{2}}{\epsilon}\cdot
	\left\{\ln(2\epsilon +1)\left[1-\frac{2(\epsilon+
	1)}{\epsilon^{2}}\right]+\frac{\epsilon+8}{2\epsilon}-
	\frac{1}{2(2\epsilon+1)^{2}}\right\}
	\label{e:kn}
\end{equation}

\noindent where $\epsilon=\hbar\omega/m_{\mathrm{e}}c^{2}$.
Therefore, in Eq.~(\ref{e:edd}), the Thompson cross section must be
replaced by Klein--Nishina formula, while all other quantities remain
unchanged. When $\epsilon >> 1$, in the ultrarelativistic case, Eq.~(\ref{e:kn}) becomes:

\begin{equation}
	\sigma_{\mathrm{KN}}=\frac{\pi r_{\mathrm{e}}^{2}}{\epsilon}(\ln2\epsilon
	+\frac{1}{2})
	\label{e:knrel}
\end{equation}

For photons produced in proton--antiproton annihilation, we have
$\epsilon \approx 391$ (for 200~MeV photon), which results in
$\sigma_{\mathrm{KN}}\approx 4.6\cdot10^{-31}$~m$^{2}$. Therefore, the
Eddington limit of matter--antimatter case:

\begin{equation}
	L_{\mathrm{am}}\approx 2\cdot 10^{40}\frac{M}{M_{\odot}}~\mathrm{erg/s}
	\label{e:eddam}
\end{equation}

Eq.~(\ref{e:eddam}) shows that in the case of accretion power from a
close encounter of a star and an antistar, the balance of radiation
pressure and gravitational force occurs at higher luminosities, owing
to smaller cross section for high--energy photons produced by
annihilation of protons and antiprotons. The new limit can be higher
when we deal with direct annihilation
$(p+\bar{p}\rightarrow\gamma+\gamma)$
or with heavier nuclei. For example, the direct annihilation of helium leads to:

\begin{equation}
	L_{\mathrm{am}}\approx 2.5\cdot 10^{41}\frac{M}{M_{\odot}}~\mathrm{erg/s}
	\label{e:eddam2}
\end{equation}

When photon energy is greater than $2m_{e}c^{2}$, pair production can
take place in the field of a nucleus. The pair production is not
possible in free space, because of the conservation of momentum and
energy. In matter and antimatter environment, nuclei annihilate among
themselves, and therefore they cannot help in the pair production.

However, the role of the nucleus in balancing energy and momentum,
can be replaced by another photon with lower energy. The
electron--positron pair can be created in a head--on collision of a
photon with energy $\epsilon_{1}$ with another photon with lower energy (see
Longair~\cite{LONGAIR}):

\begin{equation}
	\epsilon_{2}\geq \frac{m_{e}^{2}c^{4}}{\epsilon_{1}}
	\label{e:pair}
\end{equation}

For 300~MeV photons, it is necessary to have 1~keV photons, which are
present in accretion processes; therefore 300~MeV photons can be absorbed
from the environment. Higher--energy photons require a lower
energy photons, up to the Microwave Background Radiation for photons
of $4\cdot 10^{14}$~eV.

It is worth noting that also the decay of $\mu^{\pm}$ create
electron--positron pairs. These particles interact with the magnetic field
of the stars and emit syncrotron radiation (Alfv\'en~\cite{ALF2}).
The energy released cannot be larger than $m_{e}c^2/3$ and the frequency spectrum
depends on the geometry and the intensity of the magnetic field.
Moreover, $e^{+}e^{-}$ pairs can eventually generate X--rays by inverse
Compton scattering of startlight (Carlqvist and Laurent~\cite{CARL}).

\section{Collisions and merging}
As the annihilation is a surface process (a few mean free paths
thick), the release of energy due to annihilation is not large enough
to affect the kinematics of the collision (Alfv\'en~\cite{ALF2}).
Massive stars can collide and release energy of surface annihilation in a
few seconds, as a gamma--ray burst.

One of the first theory invoked to explain the origin of cosmic
gamma--ray bursts was the collision of asteroids and comets of
antimatter with stars of matter (Sofia and Van Horn~\cite{SOFIA1},
Sofia and Wilson~\cite{SOFIA2}). Here we propose another version
of that theory, specifically the collision of a star and an antistar.

There are already some theories that explain gamma--ray emission from
merging of massive stars (Eichler et al.~\cite{EICHLER}, Narayan
et al.~\cite{NARAYAN}), but they have an upper limit in the energy
released (up to $10^{53}$~erg) and they have to use beaming effects, or
other, in order to explain higher values for energy.

In the case of collision of a star with an antistar, the energy
released derives from annihilation.  Therefore, it depends simply on
the total amount of matter and antimatter, does not need of any
particular effects, and has virtually no upper limit.

A crude estimation of an order of magnitude shows that the total mass
(of matter and antimatter) required to generate a GRB of
$E=10^{54}$~erg is:

\begin{equation}
	m_{\mathrm{tot}}=\frac{E}{c^{2}}\approx 1.1\cdot
	10^{33}~\mathrm{g}
	\label{e:GRB1}
\end{equation}

\noindent that is equal to about $0.56$ solar masses; $c$ is the speed
of light in vacuum.  The collision time is of the order of some
milliseconds and this can be compatible with very short GRB (see Cline
et al.~\cite{CLINE}, Krennrich et al. ~\cite{SHORTGRB}).

Once the merging is occurred, then the new star has a
particular behaviour, because of the mixing of matter and antimatter.
Such a star has already been studied, even though with no reference to such a case,
by Unno and Fujimoto~(\cite{UNNO}). They found that a very massive star
($10^7$ solar masses or higher) made of matter and antimatter can release $10^{61}$~erg
of energy in $10^{5}-10^{6}$~yr, with a luminosity of $10^{47}$~erg/s.
These data are comparable with that of QSO.

The existence of a correlation between QSO and GRB was already invoked by
Sillanp\"a\"a~(\cite{QSO2}) and Schartel et al.~(\cite{QSO1}). In our case,
the correlation can be explained so that the GRB is the ``first light'' of a
future QSO, even though it is still unknown how this type of system can evolve.

\section{Some examples}
Steigman~(\cite{STEIG1}) wrote his paper in 1976.
Later on, Allen~(\cite{ALLEN}) analysed again the constraints on
annihilation in active galaxies and used more recent data (in 1981). Today
we have even more recent data, so we can do another evaluation of order of
magnitudes.

In his calculations for photon flux above 100~MeV,
Steigman~(\cite{STEIG1}) noted that for every erg of energy going in
electron-positron pairs, there are roughly $10^4$ photons with energy above 100~MeV.
Steigman found a relationship between the observed flux and the gamma--ray flux:

\begin{equation}
S_{\gamma} [\mathrm{photons}\: \mathrm{cm}^{-2} \mathrm{s}^{-1}] \approx 10^{4}S_{obs}
[\mathrm{erg}\: \mathrm{cm}^{-2} \mathrm{s}^{-1}]
\label{e:flux1}
\end{equation}

Allen~(\cite{ALLEN}) noted that it is better to use as reference the flux at radio
frequencies, because non thermal optical emission need not to be synchrotron
radiation (we remind that an annihilation source is also a strong synchrotron source;
see Sect.~4). Therefore, Eq.~(\ref{e:flux1}) should be changed in the more suitable:

\begin{equation}
S_{\gamma} [\mathrm{photons} \: \mathrm{cm}^{-2} \mathrm{s}^{-1}] \approx 10^{4}S_{radio}
[\mathrm{erg}\: \mathrm{cm}^{-2} \mathrm{s}^{-1}]
\label{e:flux2}
\end{equation}

We can therefore set up a new table of selected strong radio sources,
calculate the expected gamma--ray flux above 100~MeV, and
compare with data available in literature (see Tab.~\ref{examples}).

\begin{table*}[t]
\caption[]{Selected examples of sources with strong synchrotron emission and comparison with
gamma--ray flux above 100~MeV (calculated and observed).}
\label{examples}
\begin{tabular}{lcccc}
\hline
Source & $S_{radio}$ at 1~GHz$^{*}$ & $S_{\gamma}$ Calc. & $S_{\gamma}$ Obs. & Notes\\
{} & [erg cm$^{-2}$ s$^{-1}$] & [ph cm$^{-2}$ s$^{-1}$] & [ph cm$^{-2}$ s$^{-1}$] & {}\\
\hline
3C273 & $5\cdot 10^{-13}$ & $5\cdot 10^{-9}$  & $1\cdot 10^{-7}$    & a \\
Cen A & $2\cdot 10^{-11}$ & $2\cdot 10^{-7}$  & $1\cdot 10^{-7}$    & a \\
Crab  & $1\cdot 10^{-11}$ & $1\cdot 10^{-7}$  & $7\cdot 10^{-6}$    & b \\
Cyg A & $2\cdot 10^{-11}$ & $2\cdot 10^{-7}$  & $2\cdot 10^{-7}$    & a \\
M87   & $3\cdot 10^{-12}$ & $3\cdot 10^{-8}$  & $4\cdot 10^{-8}$    & a \\
Sgr A & $2\cdot 10^{-11}$ & $2\cdot 10^{-7}$  & $1\cdot 10^{-6}$    & c \\
\hline
\end{tabular}
\begin{list}{}{}
\item[$^{\mathrm{*}}$] From Zombeck~(\cite{ZOMB}).
\item[$^{\mathrm{a}}$] Fichtel et al.~(\cite{GRBC2}).
\item[$^{\mathrm{b}}$] Range 50~MeV -- 10~GeV (EGRET). Macomb and Gehrels~(\cite{GRBC1}).
\item[$^{\mathrm{c}}$] Closest $\gamma$--ray source: 2EGJ1746--2852. Macomb and Gehrels~(\cite{GRBC1}).
\end{list}
\end{table*}

It is possible to see that more recent data show some cases of interest (Cen A, Cyg A, M87), 
but it is worth noting that these sources must be studied and observed with greatest details 
before to claim that they are powered by matter--antimatter annihilation. However, these
calculation show that recent data renew the annihilation hypotesis as inner engine for certain 
sources. Perhaps future $\gamma$--ray astrophysics satellites can give more reliable constraints, 
allowing a precise choice.

\section{Conclusions}
Although the existence of stars made of antimatter is still doubtful, 
we can search for other signatures of the presence of antimatter in 
the universe. This paper showed that the Eddington luminosity in the 
presence of an antistar can be substantially higher. The high--energy 
photons created by annihilation of nuclei have a 
small cross section for their interaction with electrons and then the 
effect of radiation pressure in balancing the gravitational force is 
smaller than in standard conditions. 

However, a source with luminosity higher than the standard Eddington 
limit is not sufficient to claim for the presence of antimatter.  It 
is necessary to detect several other features, such as a strong 
synchrotron radiation, neutrinos, and high--energy photons.  We have shown 
that more recent data have, for some sources, a good agreement with the 
calculated flux from annihilation. 

These results must be taken \emph{cum grano salis} and more results 
are expected from the new generation of gamma--ray astrophysics satellites.
It is \emph{not} possible to claim for the discovery of large amount of antimatter in
the universe, but it is time to renew this type of research. 

\section{Acknowledgements}
Although they have no responsibilities in this ``crazy paper'', I wish to thank
Guido Di Cocco, Elena Pian, Diego Casadei, and Massimo Cappi for useful
discussions. This research has made use of
\emph{NASA's Astrophysics Data System Abstract Service}.

\end{document}